\documentclass[sn-mathphys-num]{sn-jnl}

\usepackage{graphicx}%
\usepackage{multirow}%
\usepackage{amsmath,amssymb,amsfonts}%
\usepackage{amsthm}%
\usepackage{mathrsfs}%
\usepackage[title]{appendix}%
\usepackage{xcolor}%
\usepackage{textcomp}%
\usepackage{manyfoot}%
\usepackage{booktabs}%
\usepackage{algorithm}%
\usepackage{algorithmicx}%
\usepackage{algpseudocode}%
\usepackage{listings}%
\usepackage{comment}
\begin{document}

\title[Article Title]{Field evaluation of a wearable instrumented headband designed for measuring head kinematics}

\author[1]{\fnm{Anu} \sur{Tripathi}}\email{tripathia@rmu.edu}

\author[2]{\fnm{Yang} \sur{Wan}}\email{yang\_wan@brown.edu}

\author[3]{\fnm{Zhiren} \sur{Zhu}}\email{zhiren@umich.edu}

\author[4]{\fnm{Furkan} \sur{Camci}}\email{camci@wisc.edu}


\author[4]{\fnm{Sheila} \sur{Turcsanyi}}\email{saturcsanyi@wisc.edu}

\author[5]{\fnm{Jeneel Pravin} \sur{Kachhadiya}}\email{jeneel.kachhadiya@wisc.edu}

\author[5]{\fnm{Mauricio} \sur{Araiza Canizales}}\email{araizacaniza@wisc.edu}

\author[6]{\fnm{Alison} \sur{Brooks}}\email{brooks@ortho.wisc.edu}

\author[2]{\fnm{Haneesh} \sur{Kesari}}\email{haneesh\_kesari@brown.edu}


\author[5]{\fnm{Joseph} \sur{Andrews}}\email{joseph.andrews@wisc.edu}

\author[7]{\fnm{Traci} \sur{Snedden}}\email{traci.snedden@cuanschutz.edu}

\author[4]{\fnm{Peter} \sur{Ferrazzano}}\email{ferrazzano@pediatrics.wisc.edu}

\author[5]{\fnm{Christian} \sur{Franck}}\email{cfranck@wisc.edu}

\author*[1]{\fnm{Rika Wright} \sur{Carlsen}}\email{carlsen@rmu.edu}

\affil[1]{\orgdiv{Department of Engineering}, \orgname{Robert Morris University}, \orgaddress{\city{Moon Township}, \state{PA}, \country{USA}}}

\affil[2]{\orgdiv{School of Engineering}, \orgname{Brown University}, \orgaddress{\city{Providence}, \state{RI}, \country{USA}}}

\affil[3]{\orgdiv{Department of Mechanical Engineering}, \orgname{University of Michigan}, \orgaddress{\city{Ann Arbor}, \state{MI}, \country{USA}}}


\affil[4]{\orgdiv{Waisman Center}, \orgname{University of Wisconsin--Madison}, \orgaddress{\city{Madison}, \state{WI}, \country{USA}}}

\affil[5]{\orgdiv{Department of Mechanical Engineering}, \orgname{University of Wisconsin--Madison}, \orgaddress{\city{Madison}, \state{WI}, \country{USA}}}

\affil[6]{\orgdiv{Department of Orthopedics and Rehabilitation}, \orgname{University of Wisconsin--Madison}, \orgaddress{\city{Madison}, \state{WI}, \country{USA}}}

\affil[7]{\orgdiv{College of Nursing}, \orgname{University of Colorado Anschutz Medical Campus}, \orgaddress{\city{Aurora}, \state{CO}, \country{USA}}}

\abstract{

\textbf{Purpose} 
To study the relationship between soccer heading and the risk of mild traumatic brain injury (mTBI), we previously developed an instrumented headband and data processing scheme to measure the angular head kinematics of soccer headers. Laboratory evaluation of the headband on an anthropomorphic test device showed good agreement with a reference sensor for soccer ball impacts to the front of the head. In this study, we evaluate the headband in measuring the full head kinematics of soccer headers in the field.   \\
\textbf{Methods}
The headband was evaluated under typical soccer heading scenarios (throw-ins, goal-kicks, and corner-kicks) on a human subject. The measured time history and peak kinematics from the headband were compared with those from an instrumented mouthpiece, which is a widely accepted method for measuring head kinematics in the field.\\
\textbf{Results}
The time history agreement (CORA scores) between the headband and the mouthpiece ranged from ‘fair’ to ‘excellent’, with the highest agreement for angular velocities (0.79 ± 0.08) and translational accelerations (0.73 ± 0.05) and lowest for angular accelerations  (0.67 ± 0.06). A Bland-Altman analysis of the peak kinematics from the headband and mouthpiece found the mean bias to be 40.9$\%$ (of the maximum mouthpiece reading) for the angular velocity, 16.6$\%$ for the translational acceleration, and -14.1$\%$ for the angular acceleration.\\
\textbf{Conclusion}
The field evaluation of the instrumented headband showed reasonable agreement with the mouthpiece for some kinematic measures and impact conditions. Future work should focus on improving the headband performance across all kinematic measures. 
}


\maketitle

\section{Introduction}\label{Introduction}

Mild traumatic brain injuries (mTBI) are estimated to affect more than 55 million people worldwide annually \cite{haarbauer2021epidemiology}. 
These injuries often result from impacts to the head, which can lead to large head accelerations and brain deformations \cite{Zhan2021, Pellman, sullivan2024pediatric}. To estimate injury risk from head impacts, accurate head kinematic data must be collected. The accuracy of head kinematic data is highly dependent on the measurement system used to collect the data, with higher accuracy obtained for wearable sensors that are tightly coupled to the skull.  Instrumented mouthguards and mouthpieces have been shown to have robust sensor-skull coupling and therefore, are widely used for head kinematic measurements \cite{Karton, Buice2018chat, Tiernan2018, Cecchi2020, patton2021, hanlon2010validation, sandmo2019evaluation, schussler2017comparison, kieffer2020two, stitt2021laboratory, rich2019development, miller2018validation, kuo2016}. However, in activities where long-term use of these devices is not feasible or preferred, alternative wearable devices are needed.  This is especially true for some sports, such as soccer, where mouthguard use is often not mandated, and many players choose not to wear mouthguards due to discomfort and communication issues \cite{acosta2024perceptions}.   

To address this need, a new instrumented headband was previously developed \cite{tripathi2025laboratory}.  The headband utilizes an array of inertial measurement units (IMUs) and an adaptive filtering method based on continuous wavelet transform to improve the head kinematics reconstruction. The headband was evaluated in the laboratory setting on an anthropomorphic test device (ATD) and was shown to accurately measure the angular head kinematics for the frontal and oblique soccer ball impacts to the head \cite{tripathi2025laboratory}.  However, laboratory settings can differ significantly from field conditions.  Factors such as differences in the head motion and sensor-skull coupling between the laboratory and field setting can have a significant effect on the performance of a wearable sensor. Therefore, it is critical to evaluate the headband in the field setting on a human subject before using the headband to collect head kinematics data across a larger cohort.

In this study, we evaluate the performance of the newly developed headband on a human subject. Field studies of wearable sensors typically assess their ability to correctly identify a head acceleration event using video verification, i.e., the specificity and sensitivity 
\cite{patton2020video,kieffer2020two, kuo2022field}. However, these studies often lack an evaluation of the measured head kinematics (peak magnitudes and time-history) in the field setting, which is crucial for estimating the resulting brain deformation and injury risk. A major challenge in the field evaluation of wearable sensors is defining an appropriate reference measurement system. High-speed video has been used as a reference in past studies \cite{wu2016vivo}; however, it is challenging to obtain an appropriate field of view while also maintaining a sufficient frame rate in the field setting. Instrumented mouthpieces have also served as a reference measurement system in previous studies and are a widely accepted method for measuring head kinematics \cite{wu2016vivo}. In this study, we have chosen to compare the head kinematics measurements from the headband with measurements from a custom instrumented mouthpiece, which has previously been validated in the laboratory \cite{miller2018validation} and deployed in field studies \cite{Filben2021,tomblin2020characterization}. 

\begin{figure}[h]
\centering
\includegraphics[width=0.9\textwidth]{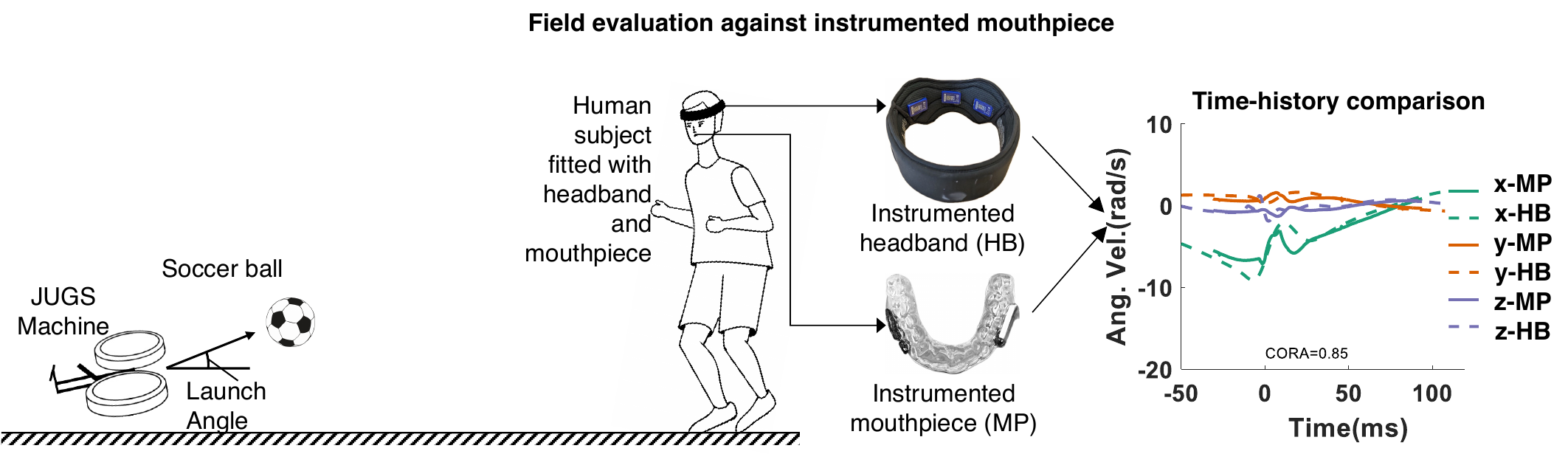}
\caption{
Field evaluation of the instrumented headband against a custom mouthpiece.  The head kinematics of a human subject were measured while performing controlled soccer headers. }\label{overview} 
\end{figure}

Figure \ref{overview} shows an overview of the experimental setup to evaluate the instrumented headband under field conditions. To enable a direct comparison of the head kinematics reconstruction from the instrumented headband and mouthpiece, a human subject performed a sequence of soccer headers (thrown-ins, goal-kicks, and corner-kicks) while simultaneously wearing both devices. 
The headband data was post-processed using the previously developed continuous wavelet-based filtering scheme \cite{tripathi2025laboratory}.  We also implemented a recently developed method to obtain the angular acceleration by solving a system of linear equations using the measurements from the sensor array, which avoids errors from numerical differentiation of the signal \cite{wan2025determiningaccelerationfieldrigid}. The resulting head kinematics (angular velocity, angular acceleration, and translational acceleration) from the headband were compared to the measurements from the mouthpiece, and the level of agreement between the two measurement systems was quantified. 

The paper is organized as follows. Section \ref{sec:method} describes the materials and methods used in the field experiments and briefly describes the headband data processing. Section \ref{Results} provides the headband kinematic results from the experiments, and Section \ref{sec:Discussion} discusses the factors affecting the performance of the instrumented headband, compares the results of this study with previous studies, and describes limitations that should be addressed in future studies.

\section{Materials and Methods}\label{sec:method}
In this section, we describe the design of the instrumented headband and mouthpiece, the field experimental setup, the data processing method for reconstructing the head kinematics, and the statistical analyses to evaluate the headband against the mouthpiece data.

\subsection{Instrumented Headband}
\label{sec:InstrumentedHeadband}
\begin{figure}[h]
\centering
\includegraphics[width=0.75\textwidth]{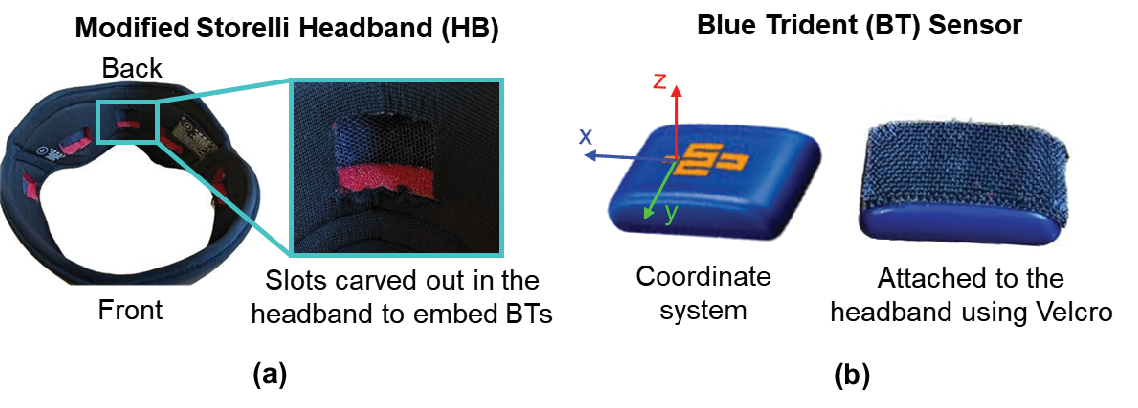}
\caption{a) Modified Storelli headband with carved-out slots in the inner surface to embed BT sensors, and b) Blue Trident (BT) inertial measurement unit (IMU) attached to the headband using Velcro}\label{HB_BT}
\end{figure}

Our wearable sensor system consists of a headband (Storelli Sports, Inc.) instrumented with five inertial measurement units (IMUs), uniformly placed from left to right around the back of the head (occipital area) (Fig. \ref{HB_BT}a). The headband design is detailed in \cite{tripathi2025laboratory} and briefly described here.
Rectangular slots of dimensions 42 $\times$ 27 $\times$ 11 mm were carved out from the inner surface of the headband, and commercially available Blue Trident IMUs (Vicon Motion Systems Limited) were placed into the slots and secured with Velcro (Fig. \ref{HB_BT}b).  

Each BT consists of a triaxial gyroscope; a low-g and a high-g triaxial accelerometer. The gyroscope measures angular velocities up to 2000 deg/sec ($\pm$ 5 deg/sec) at 1125 Hz. The low-g accelerometer measures linear accelerations up to 16 g with an accuracy of $\pm$ 0.05 g at 1125 Hz, and the high-g accelerometer can measure up to 200 g with $\pm$ 6 g accuracy at 1600 Hz. Together they provide accurate data collection over a large range of accelerations. 
All data is recorded continuously on the onboard 1 GB memory for up to 12 hours through the
Capture.U App (Vicon Motion Systems Limited).  The device does not utilize a trigger to record data.  Therefore, there is no data loss due to missed triggers (false negatives). The data processing method for reconstructing the head kinematics is described in Section \ref{sec:DataProcessing}.

\subsection{Instrumented Mouthpiece}
\label{sec:InstrumentedMouthpiece}
The instrumented mouthpiece used in this study was obtained from the Wake Forest Biomedical Engineering Team. This mouthpiece has previously been validated \cite{miller2018validation} and deployed in multiple field studies \cite{tomblin2020characterization, milef2022head}. 
The mouthpiece is custom-made based on an athlete's dental scans and is instrumented with a triaxial accelerometer and gyroscope, both with sampling frequencies of 3200 Hz. The data recording of an impact is triggered after a translational acceleration threshold of $\pm$ 3 g is exceeded for more than 3 ms. The data are recorded for 125 ms for each impact, with a duration of 31.25 ms before and 93.75 ms after the trigger. 

The mouthpiece translational acceleration and angular velocity data were post-processed by first converting the data from the sensor reference frame to the head reference frame (Fig. \ref{TimeHistories}) using subject-specific rotation matrices obtained using the athlete's dental scan. Then, the signals were filtered using a Channel Filter Class (CFC) 1000 and 155 for the translational acceleration and angular velocity, respectively. The angular acceleration was obtained by differentiating the angular velocity using a five-point stencil finite difference method using MATLAB \cite{miller2018validation}. 

\subsection{Field Experiments}
\label{sec:FieldExperiments} 

\begin{figure}[h]
\centering
\includegraphics[width=1.0\textwidth]{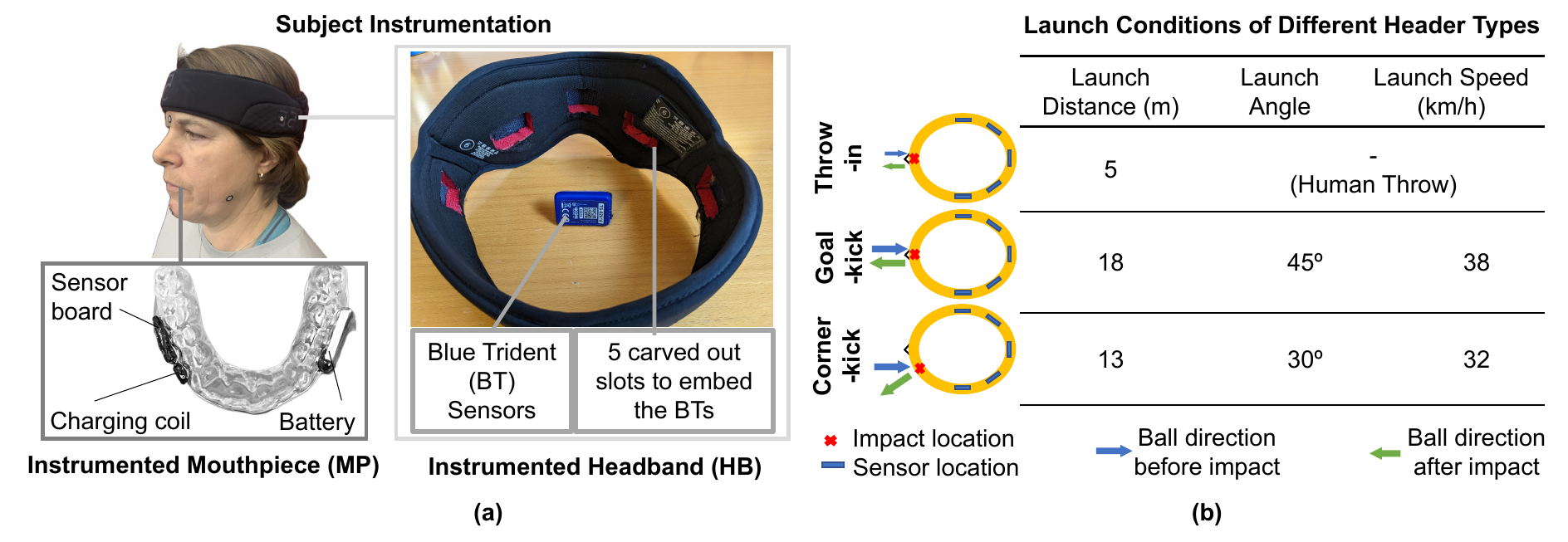}
\caption{a) Human subject simultaneously fitted with an instrumented mouthpiece and the headband, and b) launch conditions of different soccer headers used for the headband evaluation}\label{FieldSetup}
\end{figure}

The field tests evaluated the instrumented headband's performance on a human subject during a controlled header session, which was held on Feb 21, 2024. The University of Wisconsin-Madison Institutional Review Board approved the human test protocol (IRB number 2021-0048) and informed consent was obtained from the subject. The subject (adult female, former Division I player) was fitted with the instrumented headband and mouthpiece simultaneously (Fig. \ref{FieldSetup}a). A size 6 headband was selected to obtain a tight but comfortable fit, and the mouthpiece was custom-made for the athlete. 

The athlete went through a series of typical soccer headers: throw-ins, goal kicks, and corner kicks. 
For the throw-ins, another subject tossed the soccer ball towards the athlete from a distance of 5 m, and the athlete headed the ball back to the subject. 
For the goal kicks, a JUGS machine launched the ball at 38 km/h from a distance of 18 m and at a launch angle of 45\textdegree, and the athlete headed the ball back towards the JUGS machine. For the corner-kicks, the JUGS machine launched the ball at 32 km/h from a distance of 13 m and at a launch angle of 30\textdegree, and the athlete redirected the ball approximately 45\textdegree~from the incoming ball direction.
Six headers of each type were conducted, and Fig. \ref{FieldSetup}b summarizes the impact conditions for each header type.  A size 5 soccer ball with an internal air pressure of 12 psi was used in these tests. 

Each header was video-recorded (30 frames per second) with time stamps to verify the timing of the headers when comparing to the headband and mouthpiece data.  Videos were reviewed during the analysis to ensure header type correlation with the sensor data. All devices were checked for battery life and functionality before each test, and all devices remained in the recording mode for the full duration of the header session. 
Further details of the experimental design are provided in Supplementary Table S1 as per the Consensus Head Acceleration Measurement Practices (CHAMP) 2022 Reporting Guidelines for the on-field validation of wearable sensors \cite{kuo2022field}.

\subsection{Head Kinematics Reconstruction}
\label{sec:DataProcessing}

The raw kinematics data measured by the headband BT sensors were processed to reconstruct the head (skull) angular velocity, angular acceleration, and translational acceleration, and remove unwanted components of the signal caused by different sources of error. The data processing protocol is explained in detail in a previous study \cite{tripathi2025laboratory} and briefly outlined in the following sections. 

\subsubsection{Headband Angular Velocity}\label{Methods:HBAngVel}

\begin{figure}[h]
\centering
\includegraphics[width=0.6\textwidth]{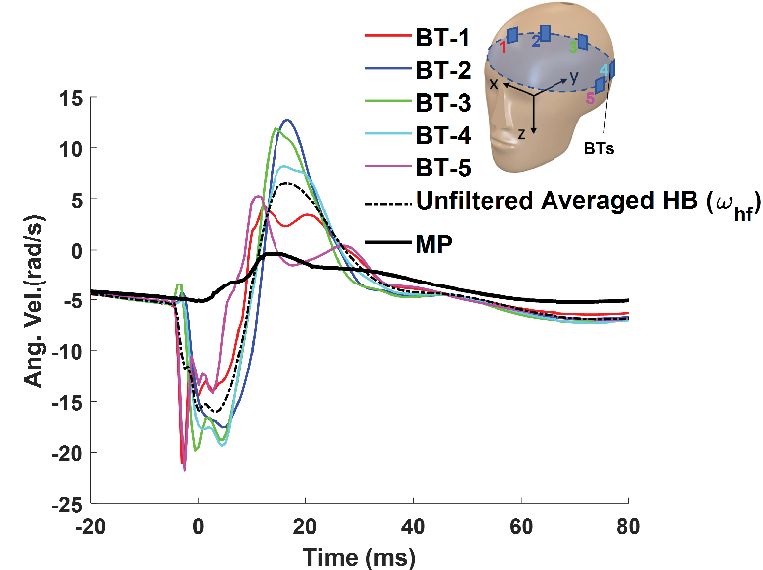}
\caption{The sagittal component ($x$-component) of the angular velocity measured by the five BT sensors in the headband is shown for a representative throw-in along with the unfiltered headband average ($\mathbf{\omega_{h}}(t)$) and the mouthpiece data (MP). All data has been converted to the head coordinate frame.  The averaged headband angular velocity ($\mathbf{\omega_{h}}(t)$) shows reduced high frequency noise compared to the individual BT data; however, it still deviates from the mouthpiece angular velocity. }\label{Fig:Averaging}
\end{figure}

The angular velocity vectors from the five BTs were first converted to the same coordinate frame as shown in Figure \ref{Fig:Averaging} and then averaged to reduce the effect of local deformations on the measured signal. The headband angular velocity $\mathbf{\omega_{h}}(t)$ is defined as

\begin{equation}
    \mathbf{\omega_{h}}(t) = \frac{ \sum_{i=1}^{N} \mathbf{\omega_i} (t)}{N} 
\end{equation}\label{Eq:avg}

\noindent where $\mathbf{\omega_i}(t)$ is the angular velocity vector time history from each BT in the head coordinate frame, and $N$ is the number of BT sensors, which is five in this study. 

After this averaging step, the high frequency noise is significantly reduced when comparing the averaged headband angular velocity $\mathbf{\omega_{h}}(t)$ to the individual BT data (BT-1 to BT-5) (Fig. \ref{Fig:Averaging}). However, we still observe a significant deviation from the mouthpiece angular velocity data. 
To further reduce the noise in the signal, a previously developed wavelet transform-based adaptive filtering method is implemented \cite{tripathi2025laboratory}.
The adaptive filtering method provides an appropriate filter cutoff frequency, $f_{0}$, to selectively remove transient noise and preserve the steady-state signal frequencies. These cutoff frequencies were uniquely acquired for each head impact since the noise content can vary based on variables such as the impact location and ball speed at the time of impact.

The adaptive filtering method is based on the continuous wavelet transform (CWT) of the headband angular velocity $\mathbf{\omega_{h}}(t)$. The CWT provides the frequency content of $\mathbf{\omega_{h}}(t)$ as a function of time or the normalized wavelet coefficients ($\bar{w}(t,\eta)$). The surface plot of the normalized wavelet coefficients is the signal scalogram. While the mouthpiece angular velocity scalogram shows minimal transient high-frequency content (Fig. \ref{Fig:Methods}a), the headband angular velocity scalogram shows significant transient high-frequency content (red box, Fig. \ref{Fig:Methods}b). The vertical line labeled $t$ = 0 ms shows the beginning of the head impact, which is defined as the time when the head linear acceleration exceeds 3 g for more than 3 ms (Figs. \ref{Fig:Methods}a,b).

Figures \ref{Fig:Methods}c-d show the normalized wavelet coefficients at two time points for the mouthpiece and headband angular velocity, respectively.
The normalized wavelet coefficient curves from the headband data at $t$ = 0 ms and $t$ = 150 ms are used to find the appropriate filtering cutoff frequency (Fig. \ref{Fig:Methods}d). 
At the time of the impact ($t$ = 0 ms), the normalized wavelet coefficients  $\bar{w}(t = 0,\eta)$ consist of the transient noise and the steady state signal (red curve).  
After a sufficiently long period, the normalized wavelet coefficients $\bar{w}(t = 150,\eta)$ consist of the steady state response only (blue curve). 
The end time point is indicated on the scalogram by the vertical line labeled $t$ = 150 ms (Fig. \ref{Fig:Methods}b). A sensitivity analysis was conducted to determine the appropriate end time point for the headband data (Supplementary Fig. S1).
Since the mouthpiece data are only available for 93 ms after the impact, we compare the normalized wavelet coefficients at 90 ms with the one at 0 ms (Fig. \ref{Fig:Methods}c), and we find significantly lower transient frequencies as compared to the headband data (Fig. \ref{Fig:Methods}d).

From the difference between the wavelet coefficients at the beginning and end time points ($\Delta \bar{w} = \bar{w}(0,\eta)  - \bar{w}(150,\eta)$), we find the frequency when the transient noise starts ($\Delta \bar{w} > 0.1$), which is denoted as $f_n$.  The highest signal frequency where $\bar{w}(t = 150\text{ ms},\eta) > 0.1$ is denoted as $f_{ss}$. The cutoff frequency, $f_{0}$, is then calculated as follows: 

\begin{equation}
    f_{0}= \begin{cases}
         \text{Max(}f_{ss},f_{n}), &   \text{Max(}f_{ss},f_{n}) < 180 \text{Hz}\\
         180 \text{ Hz}, &   \text{Max(}f_{ss},f_{n}) > 180 \text{Hz}.
        \end{cases}
\end{equation}\label{Eq:freq}

\noindent where 180 Hz is defined as the upper limit for the cutoff frequency based on SAE J211 recommendations \cite{grenke2002digital}.  The averaged angular velocity ($\mathbf{\omega_{h}}(t)$) is filtered at the cutoff frequency using a $4^{th}$ order Butterworth low-pass filter to reconstruct head angular velocity ($\mathbf{\omega_{hf}}(t)$), which is compared with the mouthpiece data (Fig. \ref{Fig:Methods}e). The wavelet-based adaptive filtering method is described in more detail in \cite{tripathi2025laboratory}.



\subsubsection{Headband Angular Acceleration}\label{Methods:HBAngAcc}


In this study, we explore two different methods for computing the angular acceleration: 1) differentiating the filtered head angular velocity using a five-point-stencil finite difference method, and 2) solving a linear system of algebraic equations using the acceleration data from multiple sensors \cite{wan2025determiningaccelerationfieldrigid}. Although numerical differentiation is widely used to compute the angular acceleration in wearable sensor studies, numerical differentiation has been shown to amplify measurement noise. The second method avoids the limitations of numerical differentiation by leveraging the sensor array in the headband.     



In the second method, we use an algorithm that has been recently developed to calculate the angular acceleration of a rigid body using three triaxial accelerometers and one triaxial gyroscope, called the A3G1-algorithm (3 accelerometers, 1 gyroscope) \cite{wan2025determiningaccelerationfieldrigid}. 
The algorithm obtains the angular and translational accelerations at a point of interest (i.e., the mouthpiece sensor location in this study) by solving the system of algebraic equations governing the translational accelerations at the locations of the three accelerometers and at the point of interest.  
The algorithm requires that the three triaxial accelerometers not lie along the same line, which is satisfied in the headband design. 

The inputs into the algorithm are the position of the point of interest, the position and principal axes directions of the three accelerometers, the translational acceleration measurements from the three accelerometers, and the angular velocity data and principal axes directions of the gyroscope. We use the filtered headband angular velocity ($\mathbf{\omega_{hf}}$) (Section \ref{Methods:HBAngVel}) as the angular velocity input, and the accelerometer data from the rightmost, the leftmost, and the back BTs. 
The algorithm solves the system of equations governing the translational accelerations ($\mathbf{a}_i(\mathbf{r}_i,t)$) at the four locations (three BT sensor locations and the mouthpiece sensor location): 

\begin{equation}
    \mathbf{a}_i(\mathbf{r}_i,t) = \mathbf{\alpha_{h}}(t) \times \mathbf{r}_i  + \mathbf{\omega_{hf}}(t) \times \mathbf{\omega_{hf}}(t) \times \mathbf{r}_i  + \mathbf{q}(t),
\end{equation}\label{Eq:a3g1}

\noindent where $\mathbf{r}_i$ is the position vector of the sensors with respect to the head coordinate frame ($i=1,2,3$ for the three BT sensors and $i=4$ for the mouthpiece), 
$\mathbf{a}_i(\mathbf{r}_i,t)$ is the translational acceleration vector in the head frame at the sensor location as a function of time, 
$\mathbf{\alpha_{h}}(t)$ is the head angular acceleration vector in the head frame, 
$\mathbf{\omega_{hf}}(t)$ is the head angular velocity vector in the head frame, 
$\mathbf{q}(t)$ is the translational acceleration  in the head frame at the origin of the head coordinate frame (i.e., where $\mathbf{r}=0$), 
and `$\times$' is the cross-product between two vectors. 

The location of each sensor was measured using vernier calipers at the beginning of the test. 
The translational accelerations at the BT locations ($\mathbf{a}_i(\mathbf{r}_i,t)$, $i=1,2,3$) were measured by the BT accelerometers, rotated to the same coordinate frame, and filtered at 260 Hz (SAE J211 recommendations based on sampling frequency). 
The rigid body motion equations were solved to obtain the head angular acceleration vector ($\mathbf{\alpha_{h}}(t)$),  the translational acceleration vector at the mouthpiece sensor location $\mathbf{a}_4(\mathbf{r}_4,t)$, and the translational acceleration vector at the origin of the head coordinate frame ($\mathbf{q}(t)$). 
The detailed mathematics of the algorithm is provided in \cite{wan2025determiningaccelerationfieldrigid}. 

\begin{figure}[h]
\centering
\includegraphics[width=1.0\textwidth]{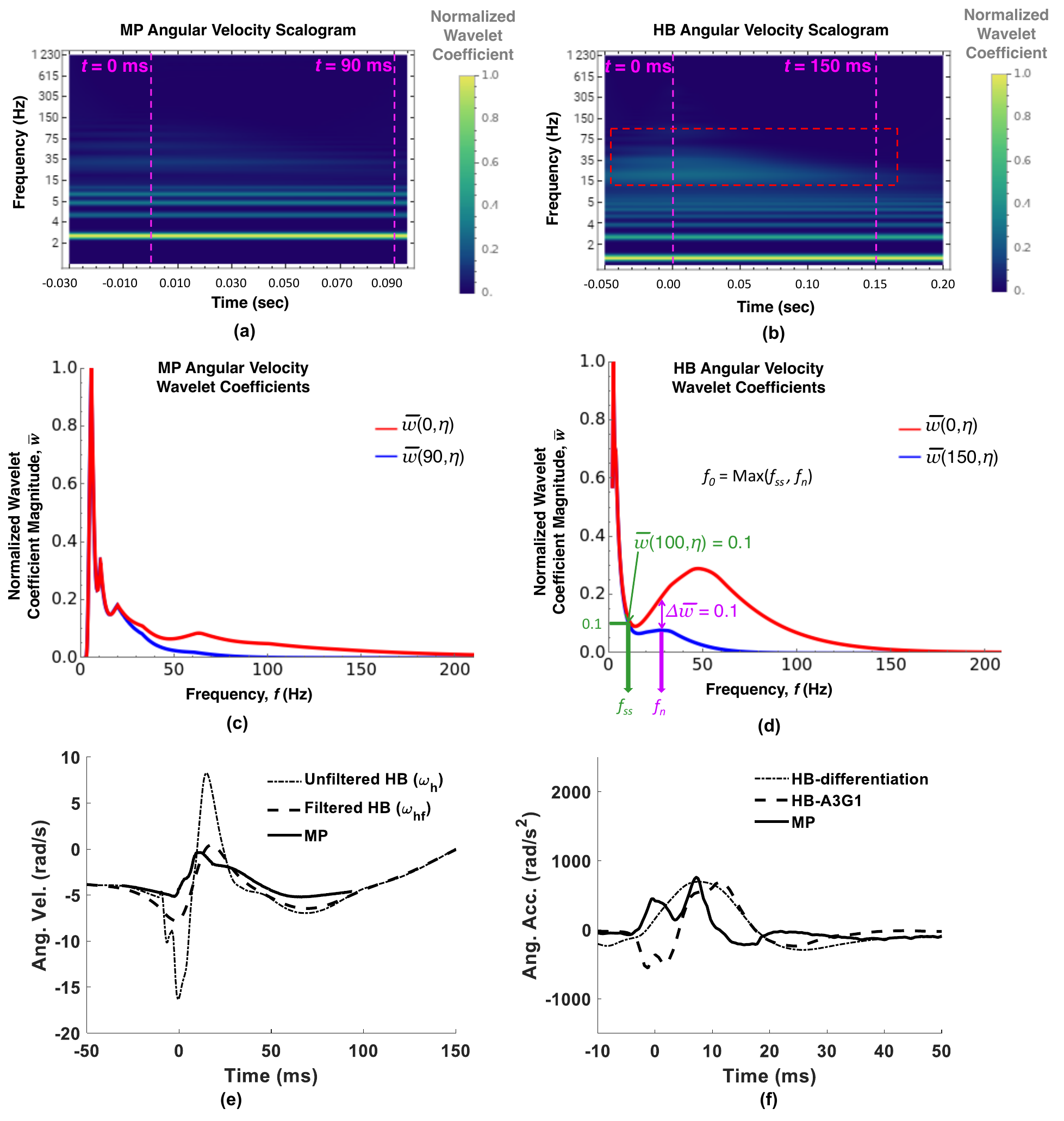}
\caption{Angular kinematics reconstruction from the headband: a) Continuous wavelet transform (CWT) scalogram of the mouthpiece (MP) angular velocity; 
b) Continuous wavelet transform  (CWT) scalogram of the headband (HB) angular velocity ($\mathbf{\omega_{h}}(t)$), with the red box highlighting the significant presence of transient frequencies in the HB angular velocities; 
c) Normalized CWT coefficients for the MP angular velocity at the beginning ($t=0$ ms) and at $t=90$ ms of the impact; 
d) Normalized CWT coefficients for the HB angular velocity at the beginning ($t=0$ ms) and at $t=150$ ms to obtain the filtering cutoff ($f_0$) based on the transient and steady state frequency content;  
e) Comparison of the unfiltered and filtered HB angular velocities (i.e., skull angular velocity reconstruction) against MP measurements;
f) Comparison of the HB angular acceleration reconstruction using the differentiation and the A3G1 algorithm methods against the MP angular acceleration obtained using differentiation.}\label{Fig:Methods}
\end{figure}

We compare the angular accelerations from the above two methods with the mouthpiece measurements to evaluate the headband performance (Fig. \ref{Fig:Methods}f). 

\subsubsection{Headband Translational Acceleration}\label{Methods:HBTransAcc}

The translational acceleration was computed from the headband data using the A3G1 algorithm, which leverages the measured data from the sensor array.  We chose to compute the translational acceleration at the location of the mouthpiece sensor ($\mathbf{a}_4(\mathbf{r}_4,t)$) to directly compare the translational acceleration from the headband with the mouthpiece measurements.
The mouthpiece, similar to several other existing wearable sensors \cite{Cecchi2020,kieffer2020two,liu2020validation}, uses a single triaxial accelerometer and a gyroscope. Therefore, it measures the translational acceleration at the location of the accelerometer. To obtain the acceleration at any other location, the rigid body motion equation (Eq. \ref{Eq:a3g1}) must be applied, 
which requires differentiation of the gyroscope data, introducing more noise. Therefore, to directly compare the translational acceleration from the headband to the mouthpiece, the translational acceleration was evaluated at the mouthpiece sensor location for both devices. 


\subsection{Statistical Data Analysis}

We performed statistical analyses to compare the skull kinematics reconstruction from the headband sensors with the mouthpiece measurements. A CORA analysis was conducted to quantify the agreement in kinematic data time histories using recommended CORA parameters \cite{giordano2016development}. The analysis quantifies the agreement between two curves based on their time-alignment (phase), peak magnitudes, and the overall shape, with equal weights applied to the three criteria. 
The analysis was performed for the angular velocity, the translational acceleration, and the angular acceleration using the entire time history available from the mouthpiece data, i.e., from 31.25 ms before the trigger to 93.75 ms afterwards. 
The time-agreement was rated using the bio-fidelity ratings defined by the CORA scores: `excellent' (CORA $>0.86$), `good' ($0.66-0.86$), `fair' ($0.44-0.65$), `marginal' ($0.26-0.44$), or `unacceptable' ($<0.26$) \cite{giordano2016development}.

We also performed a Bland-Altman analysis on the headband and the mouthpiece data for the peak angular (rotational) velocities (PRV), peak angular (rotational) accelerations (PRA), and peak translational (linear) accelerations (PLA). This analysis shows if the two methods can be used interchangeably without assuming one method to be the ground truth. The bias in the data was obtained by subtracting the peak kinematics of the mouthpiece measurements from the headband reconstructed peak kinematics for each header. 
The average of this difference provided the mean bias, and its mean $\pm$ $1.96\times$standard deviation (SD) provided the limits of agreement, within which $95\%$ of the measurements are expected to fall. A low mean and SD of the bias would indicate that the two methods can be used interchangeably. To obtain a measure of relative deviation and to compare the results with other studies, the normalized biases of each kinematic variable (PRV, PRA, PLA) were also obtained by normalizing the bias with respect to the maximum mouthpiece kinematic value (Fig. \ref{CORA_scores}b).  We also performed a paired \textit{t}-test to check the statistical significance of trends observed in the CORA and Bland-Altman analyses. Trends with \textit{p}-values smaller than 0.05 are considered statistically significant.

\section{Results}\label{Results}
\label{sec:Results}

We conducted field testing of the current headband on a human subject under typical soccer heading scenarios. The athlete performed a series of 18 controlled headers in response to throw-ins, goal-kicks, and corner-kicks (6 headers of each type). 
The reconstructed head kinematics from the headband were compared to measurements from the Wake Forest instrumented mouthpiece (MP) \cite{miller2018validation}, which has been used in numerous field studies \cite{tomblin2020characterization, milef2022head}. 
The kinematics data for all 18 impacts were recorded on both the mouthpiece and the headband sensors. The peak angular velocities (PRV), peak angular accelerations (PRA), and peak linear accelerations (PLA) measured by the instrumented mouthpiece ranged from $4-10$ rad/s, $610-3200$ rad/s$^2$, and $120-340$ m/s$^2$, respectively, for all the headers. 

\begin{figure}[h]
\centering
\includegraphics[width=1.0\textwidth]{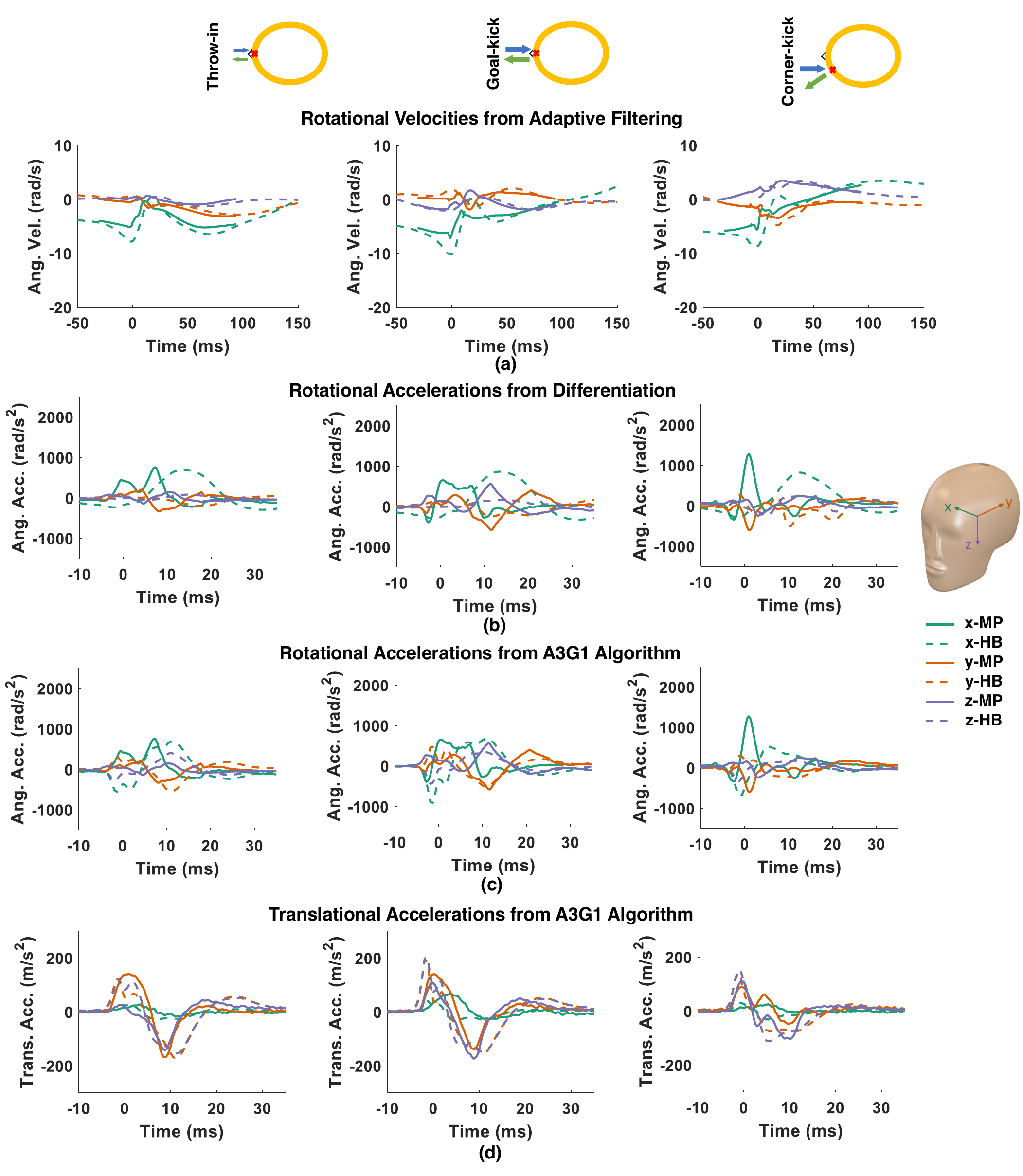}
\caption{Comparison between the mouthpiece (MP) and headband (HB) reconstruction of the a) angular velocity, b) angular acceleration from differentiation, c) angular acceleration from the A3G1 algorithm, and d) translational acceleration time histories for representative soccer headers of each type.  The time $t = 0$ ms corresponds to the time where the head linear acceleration exceeds 3 g for more than 3 ms. }\label{TimeHistories}
\end{figure}

\begin{figure}[h]
\centering
\includegraphics[width=1.0\textwidth]{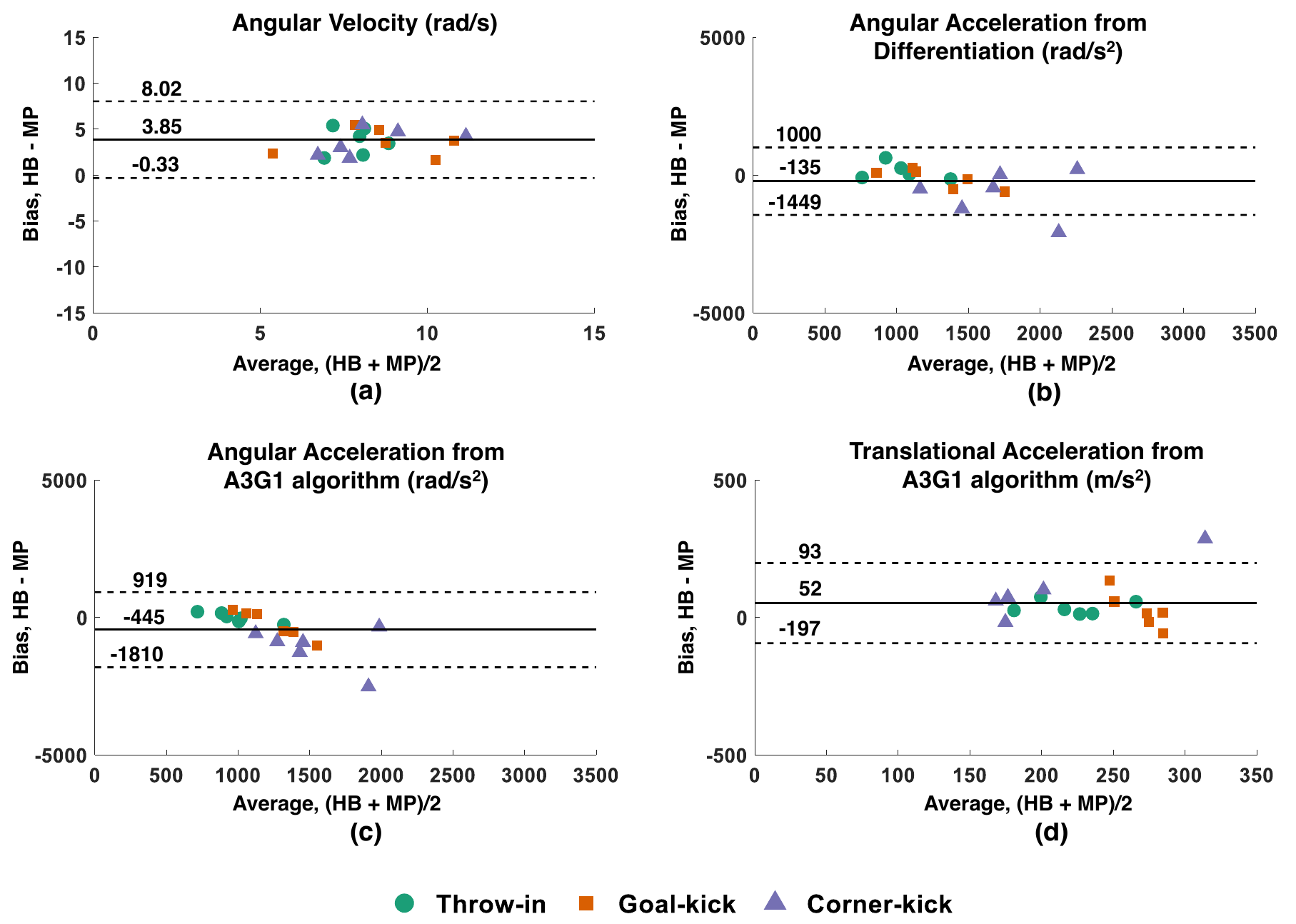}
\caption{Bland-Altman analysis showing the mean bias and the limits of agreement (mean$\pm1.96\times$SD) between the mouthpiece (MP) and headband (HB) peak kinematics for a) angular velocities, b) angular accelerations from differentiation, c) angular accelerations from the A3G1 algorithm, and d) translational accelerations from the A3G1 algorithm.}\label{BlandAltman}
\end{figure}

We find that the time-history agreement between the two devices lies in the `good' to `excellent' range for the angular velocity, with average $\pm$ standard deviation CORA scores of 0.79 $\pm$ 0.08. Representative plots of the throw-in, goal-kick, and corner-kick angular velocity time histories are shown in Fig. \ref{TimeHistories}a. 
The time-history agreement for the three header types was found to be in a similar range, with CORA scores of $0.79 \pm 0.07$ for the throw-ins, $0.79 \pm 0.07$ for the goal-kicks, and $0.79 \pm 0.10$ for the corner-kicks (Fig. \ref{CORA_scores}a). 
From the Bland-Altman analysis, we find that the mean bias between the headband and the mouthpiece PRV was 3.85 rad/s with a SD of 2.1 rad/s. The positive value of the mean bias indicates that the headband over-predicted the peak angular velocities when compared to the mouthpiece. No proportional bias exists between the PRVs from the two devices (Figs. \ref{BlandAltman}a, \ref{CORA_scores}b).

For the head angular accelerations obtained from the differentiation method, the time-history agreement between the two devices lies in the `fair’ to `good' range (CORA = 0.62 ± 0.08). Representative plots of the throw-in, goal-kick, and corner-kick angular acceleration time histories are shown in Fig. \ref{TimeHistories}b. The time-history agreement lies in a similar range for the throw-ins (CORA = 0.62 $\pm$ 0.10), goal kicks (0.63 $\pm$ 0.09), and corner kicks (0.62 $\pm$ 0.07)  (Fig. \ref{CORA_scores}a). 
The bias between the headband and the mouthpiece PRA had a mean of -135.2 rad/s$^2$ and a SD of 358.0 rad/$s^2$. The negative mean bias indicates that the headband under-predicted the peak angular accelerations compared to the mouthpiece on average. We observe the magnitude of the bias (absolute value) to increase with increasing PRA values (Fig. \ref{BlandAltman}b). The bias was also dependent on the header type, with the throw-ins showing positive mean bias, and the goal-kicks and corner-kicks showing negative mean biases. The corner-kicks have the largest bias magnitudes compared to throw-ins and goal-kicks (Fig. \ref{CORA_scores}b).

For the head angular accelerations calculated algebraically (`the A3G1 algorithm'), the time-history agreement between the two devices lies in the `fair’ to `good' range (CORA = 0.67 ± 0.06).  The CORA scores were higher for the A3G1 method compared to the differentiation method (CORA = 0.62 ± 0.08). Representative plots of the throw-in, goal-kick, and corner-kick angular acceleration time histories are shown in Fig. \ref{TimeHistories}c. The time-history agreement lies in a similar range for the throw-ins (0.67 $\pm$ 0.04), the goal-kicks (0.68 $\pm$ 0.09) and the corner-kicks (0.66 $\pm$ 0.06) (Fig. \ref{CORA_scores}a). Note that the mouthpiece angular acceleration time histories shown in Fig. \ref{TimeHistories}c were obtained via numerical differentiation. The A3G1 algorithm was able to capture the tail end of the acceleration curve (after $\sim$15 ms from the initial time of the impact) reasonably well for some impacts, 
but could not capture the peaks occurring before 15 ms (Fig. \ref{TimeHistories}c, Supplementary Fig. S4). 
The Bland-Altman analysis of the PRA obtained from the A3G1 method had a higher mean bias of -445.0 rad/s$^2$ and a SD of 696.3 rad/$s^2$ compared to the differentiation method. Similar to the differentiation method, the magnitude of the bias (absolute value) increases with PRA values, with the headband generally under-predicting the PRAs compared to the mouthpiece (Fig. \ref{BlandAltman}c). Since corner kicks experienced the highest PRAs, they had the highest bias between the headband and the mouthpiece of all header types (Figs. \ref{BlandAltman}c, \ref{CORA_scores}b). 

The linear accelerations at the mouthpiece location were also obtained from the headband using the A3G1 algorithm and evaluated against the mouthpiece data. 
The time-history agreement for the linear accelerations for all header types was found to be in the `good' range (CORA = 0.73 ± 0.05). Representative plots of the throw-in, goal-kick, and corner-kick linear accelerations are shown in Fig. \ref{TimeHistories}d. The time-history agreement was similar between throw-ins (CORA = 0.72 $\pm$ 0.06), goal kicks (0.76 $\pm$ 0.03), and corner kicks (0.72 $\pm$ 0.04) (Fig. \ref{CORA_scores}a). The mean bias between the headband and the mouthpiece peak linear accelerations (PLA) from the Bland-Altman analysis was 52 m/s$^2$, and the SD was 74.3 m/s$^2$. The positive mean bias shows that the headband over-predicted the PLA compared to the mouthpiece on average, and no proportional bias was found (Fig. \ref{BlandAltman}d). The bias for corner-kicks was higher than for throw-ins and goal-kicks (Fig. \ref{CORA_scores}b).

\section{Discussion}
\label{sec:Discussion}

This study aims to evaluate an alternative wearable device to mouthguards for head kinematics measurements in soccer, where repeated heading of soccer balls has raised concerns of long-term risk to the brain \cite{Ling2017,Hales2307}.  While mouthguards and mouthpieces are widely used for head kinematics measurements due to their robust sensor-skull coupling, having alternate wearable sensor options may lead to higher athlete compliance in large-cohort soccer studies.  This study focuses on a previously developed instrumented headband, which was designed to be user-friendly and also improve upon existing headband designs by incorporating an array of IMU sensors and implementing a new adaptive filtering method \cite{tripathi2025laboratory}.  The headband was previously shown to accurately measure the angular head kinematics for soccer ball impacts to the front of the head in the laboratory setting \cite{tripathi2025laboratory}.  This study extends this prior work by evaluating the headband on a human subject under more realistic field conditions.

The previous laboratory tests   \cite{tripathi2025laboratory} differed from typical field conditions in several major ways.  In the laboratory tests, the headband was tested on a Hybrid-III headform, which responds passively when impacted with a soccer ball.  In a realistic soccer heading scenario, an athlete actively accelerates their head toward the ball, resulting in significant pre-impact kinematics which was absent in the laboratory data.  The presence of hair and soft tissue at the head-headband interface, which was absent in the laboratory tests, can affect the sensor-skull coupling. 
The stiffness of the human neck and skull is also different between the Hybrid-III ATD and \textit{in-vivo} conditions  \cite{thompson2022review}. These differences can affect the contact duration with the soccer ball, thereby altering the noise duration, noise frequencies, and the data-processing requirements. Given these differences in testing conditions, it is critical to evaluate the headband in the field setting on a human athlete before implementing the device in future head kinematics studies.

In this study, the headband was evaluated on a human athlete for typical soccer heading scenarios (thrown-ins, goal-kicks, and corner-kicks).  The peak magnitude and time histories of the measured angular velocity, angular acceleration, and translational acceleration from the headband were compared with measurements from a custom instrumented mouthpiece, which had been previously validated in the laboratory  \cite{miller2018validation} and deployed in multiple field studies characterizing head impacts in soccer \cite{tomblin2020characterization, Filben2021}.  

In the following sections, we discuss the head kinematics
agreement between the headband and mouthpiece data, and we compare the performance of the headband in the laboratory and field settings. We also compare the results of this study with other \textit{in vivo} studies of wearable sensors.  Finally, we discuss the limitations of this study and provide suggestions for future improvement.



\subsection{Angular Velocity}\label{Diss:AngVel}

The head angular velocity was measured using a gyroscope in both the instrumented headband and the mouthpiece. The mouthpiece angular velocity was sampled at 3200 Hz and filtered using CFC 155, as per its data-processing recommendations \cite{rich2019development}.
The headband angular velocity was averaged from the five IMU sensors (1125 Hz sampling frequency) and adaptively filtered to remove the transient noise while preserving the steady-state frequency content (Section \ref{Methods:HBAngVel}). 

\begin{figure}[h]
\centering
\includegraphics[width=0.9\textwidth]{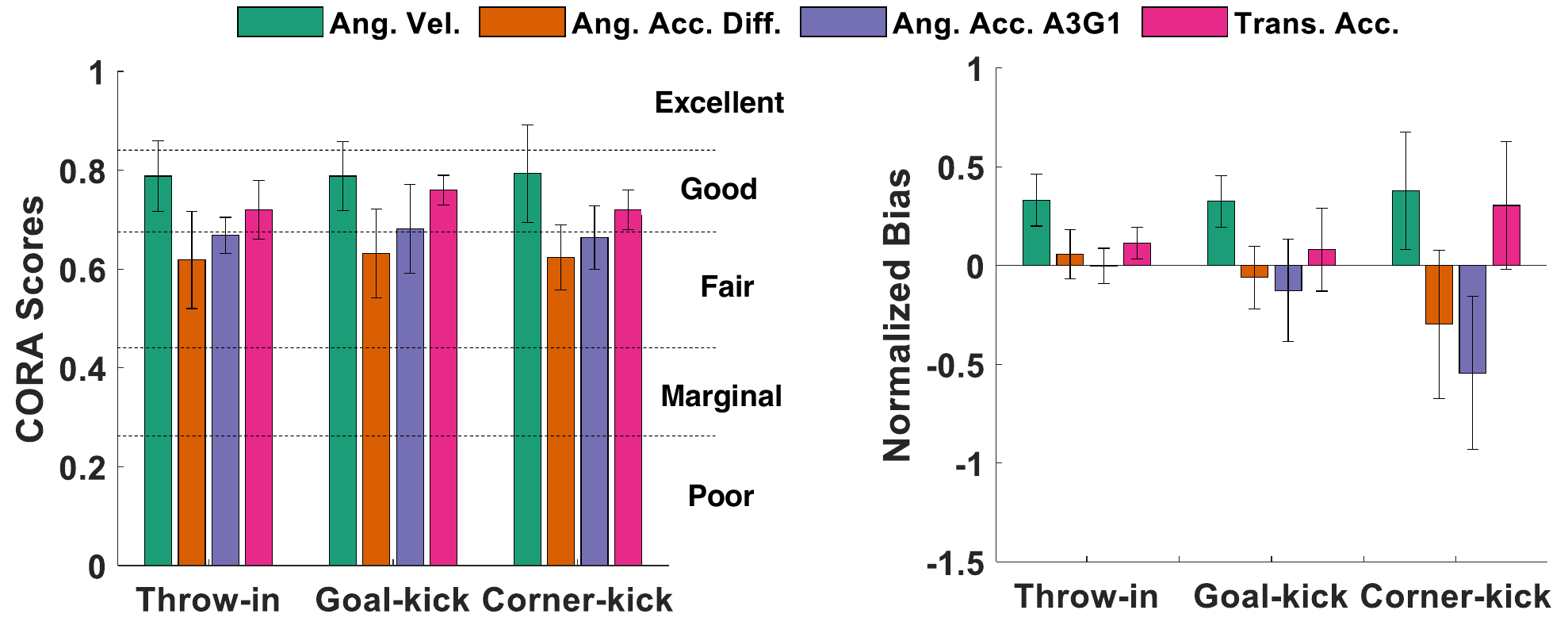}
\caption{Agreement between the mouthpiece (MP) and headband (HB) kinematics reconstruction for the three types of soccer headers (throw-ins, goal-kicks, and corner-kicks). a) The CORA scores show the time-history agreement. b) The normalized bias from the Bland-Altman analyses, (HB-MP)/max(MP), shows the agreement in the peak kinematics reconstruction (the bars represent the average values, and the error bars represent the standard deviation). }\label{CORA_scores}
\end{figure}

Overall, the angular velocity reconstruction from the headband resulted in a `good' to `excellent' time history match (Average CORA = 0.79 $\pm$ 0.08) with the mouthpiece data for all header types, as shown in Fig. \ref{CORA_scores}a. 
In the laboratory tests \cite{tripathi2025laboratory}, we had found similar time-history agreement between the headband and the reference sensor (DTS 6DX PRO-A sensor from Diversified Technical Systems, Inc.) for frontal and front-oblique impacts (Average CORA = 0.83 $\pm$ 0.06). 
However, the normalized mean bias (with respect to the peak value of the reference sensor) from the Bland-Altman analysis of the peak rotational velocity (PRV) was much lower in the laboratory tests (Normalized Bias = 9.0 $\pm$ 9.0\%) than in the field tests (Normalized Bias = $40.9 \pm 22.6\%$). A lower normalized mean bias is desirable. The normalized bias results from the field tests are shown in Fig. \ref{CORA_scores}b, and the results from the laboratory tests can be found in Tripathi et al. \cite{tripathi2025laboratory}.   %

There are several other differences in the angular velocity time histories between the laboratory and field data. The peak angular velocities in the field data occurred right before or near the time of impact with the soccer ball (within 5 ms).  This was due to the pre-impact head motion (Fig. \ref{TimeHistories}a). In contrast, the peak angular velocity in the laboratory impacts on a Hybrid-III headform occurred $20 - 25$ ms after the initial impact with the soccer ball.  As a result of this delayed kinematics, the transient noise from the impact had time to dissipate before the peak kinematics was reached, which enabled the noise to be more effectively filtered in the laboratory tests.  The transient noise coincided with the peak head motion in the field tests, making the assessment of the PRV more challenging.  In fact, we see the highest deviations between the headband and the mouthpiece data in the field tests near the beginning of the impact (Fig. \ref{TimeHistories}a).  

The difference between the laboratory and field conditions affected the efficacy of the adaptive filtering. The average cutoff frequency obtained from the adaptive filter was lower in the field tests ($56 \pm 33$ Hz) compared to the laboratory tests ($126\pm 50$ Hz). The lower average cutoff frequency for the field data resulted in more severe filtering, which can result in signal loss in the field headband data \cite{tripathi2025laboratory,abrams2024biofidelity}.  
We also see short duration pulse features in the mouthpiece data at the beginning of the impact (Supplementary Fig. S1, red circle), which become more prominent as the ball velocity is increased. These transient features are indistinguishable from the noise in the wavelet analysis of the headband data, and therefore, are not present in the filtered headband angular velocity.

The results of this study reveal some of the limitations of the adaptive filtering method. The most significant being the overlap in the signal and noise frequencies, which in the field tests limits the efficacy of the adaptive filtering method and leads to deviations in the peak angular velocities from the two devices. Reducing the overlap in these frequencies through a re-design of the headband should lead to an improvement in angular velocity reconstruction, which is further discussed in Section \ref{Diss:lim}. These differences in the results emphasize the importance of evaluating the headband on human subjects under realistic field conditions.



\subsection{Angular Accelerations}\label{Diss:AngAcc}

In this section, we compare the differentiation and the algebraic reconstruction (A3G1 algorithm) methods for obtaining an improved estimate of the angular acceleration from the instrumented headband.

The first method (differentiation method) was applied to both the mouthpiece and headband data. In the differentiation method, the angular acceleration was obtained using a five-point stencil finite difference of the angular velocity. This method was also applied to obtain the angular accelerations from the headband and reference sensor data in the previous laboratory tests \cite{tripathi2025laboratory}. 
The second method (A3G1 algorithm) uses data from three accelerometers and one gyroscope.  This method was only applied to the headband field data to algebraically reconstruct the angular and translational accelerations, as described in Section \ref{Methods:HBAngAcc}.

The angular acceleration obtained from the A3G1 method (CORA score = $0.67\pm0.06$) resulted in improved time-history agreement with the mouthpiece angular acceleration data compared to the differentiation method (CORA score = $0.62\pm0.08$) (Fig. \ref{TimeHistories}b,c).  However, the bias was $-135.2\pm358$ rad/s$^2$ ($-4.27\pm11.3\%$) for the differentiation method compared to a bias of $-445.2\pm696.3$ rad/s$^2$ ($-14.1\pm21.9\%$) for the A3G1 method, indicating that the peak angular acceleration (PRA) agreement was better for the differentiation method  (Fig. \ref{CORA_scores}b). Both methods resulted in the headband generally under-predicting the angular accelerations as compared to the mouthpiece (which used the differentiation method) ($\textit{p}<0.05$).  Both methods also revealed a proportional bias, where the bias increased with increasing PRA magnitudes (Figs. \ref{BlandAltman}b, c).  The bias was also dependent on the header type (Fig. \ref{CORA_scores}b).  
The peak angular acceleration from the A3G1 method resulted in a lower bias (higher agreement with the mouthpiece results) for the low-intensity throw-in headers, but the bias was higher for the goal-kicks and corner-kicks. However, these differences between the two methods were not statistically significant for any header type ($\textit{p}>0.05$)  (Fig. \ref{CORA_scores}a).

Since angular acceleration is the time derivative of angular velocity, 
the angular acceleration results obtained via differentiation are very sensitive to the high-frequency content in the angular velocity data. 
The mouthpiece angular velocity was filtered at a higher cutoff frequency (CFC 155) and retains some transient high-frequency content after filtering with a low-pass filter (scalograms in Fig. \ref{Fig:Methods}b). 
A much lower average cutoff frequency ($56 \pm 33$ Hz) was applied to the headband data, resulting in more severe filtering of the high-frequency content and lower agreement between the mouthpiece and the headband (Fig. \ref{TimeHistories}b). 
The A3G1 method avoids differentiation, thereby preventing the amplification of high-frequency errors in the headband angular velocity reconstruction.
However, since the A3G1 algorithm uses angular velocity and translational acceleration measurements from multiple sensors, any errors in these signals will affect the accuracy of the angular acceleration results.  Therefore, the A3G1 algorithm did not yield substantial improvements overall (Fig. \ref{TimeHistories}c). 

When comparing the angular acceleration results of the field tests to the previous laboratory tests \cite{tripathi2025laboratory}, we found the normalized bias of the field tests for both methods (differentiation method = $-4.27\pm11.3\%$; A3G1 method = $-14.1\pm21.9\%$ of maximum mouthpiece PRA) to be comparable in magnitude to the laboratory test results for the front and front-oblique impacts (bias = $7 \pm 20\%$ of the maximum reference sensor PRA). 


\subsection{Translational Accelerations}\label{Diss:TransAcc}


The time-history agreement between the headband and mouthpiece translational acceleration data was within the `good' range for all headers (CORA = 0.73 $\pm$ 0.05).  The Bland-Altman analysis showed a small bias of 52$ \pm $74 m/s$^2$ ($16.6 \pm 23.7\%$), which provides confidence in the algebraic kinematics reconstruction method for the translational acceleration. 
The translational acceleration results from the A3G1 algorithm showed no proportional bias between the two devices, similar to the angular velocity (Fig. \ref{BlandAltman}d).  However, the normalized mean bias was in general higher for the corner kicks compared to the other types of headers (Fig. \ref{CORA_scores}b), so caution must be taken when reconstructing the translational acceleration for certain types of headers.

 
\subsection{Comparison with other wearable devices}\label{Diss:Comp}

The evaluation of wearable devices under realistic field conditions poses a challenge, especially with defining an appropriate reference measurement system.  Instrumented mouthguards or mouthpieces have been used as a reference measurement system in several studies, which have evaluated skin-mounted \cite{wu2016vivo}, helmet-mounted \cite{holcomb2024field}, and skull cap-mounted \cite{wu2016vivo} wearable sensors on human subjects.  Here, we compare the performance of the instrumented headband from this study with other wearable sensor systems that have been evaluated on a human subject while heading a soccer ball.

Wu et al. \cite{wu2016vivo} compared the head kinematics measurements from a skull cap and a skin patch against the measurements from an instrumented mouthguard on an adult male for 10 frontal impacts with a soccer ball. 
The range of the measured mouthguard kinematics was 750 $\pm$ 300 rad/s$^2$ (angular acceleration) and 91.2 $\pm$ 18.6 m/s$^2$ (linear acceleration), making these impacts milder compared to this study ($1450 \pm 650$ rad/s$^2$ and $200 \pm 50$ m/s$^2$).  
The Wu study proposed a spring-dashpot model of the sensor-skull system to account for the skull cap and skin patch errors relative to the mouthguard. However, the models only accounted for the sagittal rotation and anterior-posterior translation. 

The peak kinematic performance of the devices was quantified by measuring the average peak difference with the mouthguard (mean bias).  The Wu study found a bias of 10.3 $\pm$ 8 rad/s, 4300 $\pm$ 2700 rad/s$^2$, and 491 $\pm$ 304 m/s$^2$ for the skull cap; and a bias of 9.9 $\pm$ 4 rad/s, 2500 $\pm$ 1200 rad/s$^2$, and 147.2 $\pm$ 68.7 m/s$^2$ for the skin patch for the angular velocity, angular acceleration, and translational acceleration, respectively. Both the skull cap and the skin patch over-predicted the mouthpiece for all three kinematic measures. In comparison, our instrumented headband had smaller biases of 3.85 $\pm$ 2.13 rad/s, -445 $\pm$ 696.2 rad/s$^2$ (A3G1 method), and 52 $\pm$ 74.3 m/s$^2$, with the headband over-predicting PRV and PLA and under-predicting PRA. The Wu study did not provide peak kinematics evaluation after model fitting.

The time history agreement in the Wu study was reported using the normalized root mean square (NRMS) error over 24.4 ms centered around the peak value (25 data points). The NRMS error was normalized using the peak kinematic value from the mouthguard for each kinematic measure.  The NRMS error for the skull cap was found to be 110 $\pm$ 80$\%$, 500 $\pm$ 40 $\%$, and 230 $\pm$ 260$\%$; and the NRMS error for the skin patch was 150 $\pm$ 76$\%$, 290 $\pm$ 230 $\%$, and 120 $\pm$ 40$\%$ 
for the angular velocities, angular accelerations, and translational accelerations, respectively. 
After accounting for the error through the spring-dashpot model, the NRMS time-history agreement for the anterior-posterior translation improved to 20.4 $\pm$ 11$\%$ for skin patch and 31.1 $\pm$ 11$\%$ for the skull cap. The sagittal angular velocity was modeled only for the skin patch, which resulted in an NRMS error of 20 $\pm$ 8$\%$.

In comparison, our headband provided significant improvement in time-history agreement as calculated over 24.4 ms centered around the peaks at the beginning of the impact (40 data points).  
We found the NRMS (RMS) errors to be -7.4 $\pm$ 8.5$\%$ (-0.70 $\pm$ 0.80 rad/s) for the angular velocities, 2.3 $\pm$ 2.4$\%$ (104.8 $\pm$ 107.4 rad/s$^2$) for the angular accelerations, and 3.6 $\pm$ 1.4$\%$ (11.47 $\pm$ 4.4 m/s$^2$) for the translational accelerations.  Given the lack of \textit{in-vivo} evaluation of wearable sensors for head kinematics measurements under realistic soccer heading conditions, there are limited studies to serve as a comparison basis; however, we find that the instrumented headband has improved results compared to the skin patch and skull cap that were assessed in Wu et al. \cite{wu2016vivo}.  Despite these improvements, further work is needed to bring the headband performance closer to the performance of instrumented mouthpieces.

\subsection{Limitations and Future Work}
\label{Diss:lim}

There are several limitations of this study.  First of all, the sample size was small.  The instrumented headband was evaluated on one human subject for 18 total headers (6 of each type).  Future studies should be conducted on a larger number of subjects and head impacts.  The instrumented headband should also be evaluated over a wider range of ball impact speeds and for different types of impacts, such as ground-to-player, player-to-player, and simultaneous impacts, to assess the suitability of the headband for different on-field impact conditions. The effect of headband fit on the performance on the headband and the slipping potential of the headband under high impact loads was not assessed in this study and should  be evaluated in the future.

Another limitation of this study is the use of the mouthpieces as the reference measurement system.  While mouthguards and mouthpieces are the most widely accepted method for measuring head kinematics under realistic field conditions, they do not serve as a ground truth measurement.  Some studies have shown errors in mouthguard measurements due to mandible motion, especially in higher magnitude impacts \cite{abrams2024biofidelity,kuo2016}. 
The mouthpiece used in this study has  been evaluated on a clenched mandible ATD but has not been evaluated on a cadaver \cite{rich2019development}. The rigid retainer form factor of the mouthpiece has been found to provide tighter coupling with the upper dentition compared to most mouthguards \cite{miller2018validation}; however, further studies are needed to evaluate the robustness of the coupling under a wide range of conditions.  
Given the possibility of errors with mouthpiece measurements, future validation studies should include additional reference measurements, such as 3D high-speed video, to further verify the results.


This study also highlights the limitations of the wavelet-based adaptive filtering in field conditions due to the overlap of the signal and noise frequencies. Therefore, future efforts should be directed towards reducing the overlap between the signal and noise frequencies and improve the signal-to-noise ratio. 
The previous laboratory tests showed that the transient noise arises mainly due to the loading itself and is isolated to the headband and the surrounding materials \cite{tripathi2025laboratory}. Since the headband material stiffness and damping affect the noise frequencies, the headband material can be chosen to better separate the noise frequencies from the signal. This would lead to more efficient filtering and better reconstruction of head kinematics. 
Future algorithm development should also focus on quantifying the confidence interval for each reconstruction based on the overlap of the signal and noise frequencies and the signal-to-noise ratio. 

Other headband designs can also be explored that minimize the potential of headband slipping under high-magnitude impacts. 
Also, several advanced reconstruction algorithms have been developed that provide improved kinematic reconstruction of noisy data \cite{wan2022determining,rahaman2020accelerometer}, but require a non-coplanar arrangement of IMUs \cite{wan2022determining,rahaman2020accelerometer}. Therefore, headband designs that allow non-coplanar sensor arrangements
while also maintaining player comfort over the course of a practice or game should also be explored. 
Finally, selecting sensors with higher sampling rates would also improve the time resolution of the captured data.  
With these improvements, the performance of instrumented headbands will be brought closer to the performance of instrumented mouthpieces, providing an alternative wearable device for measuring head kinematics.



\backmatter

\bmhead{Acknowledgments}
The authors gratefully acknowledge Wake Forest Biomedical Engineering Team for providing the mouthpieces and post-processing codes. We are also grateful to Prof. Andrew Watson, Allison Schwartz, Mike Powers, and Jessica Park for helping with the experiments. We also acknowledge Alice Lux Fawzi and the PANTHER program for facilitating fruitful discussions and collaborations.

\section*{Declarations}

\begin{itemize}
\item \textbf{Funding:} The authors gratefully acknowledge the support from the University of Wisconsin-Madison Office of the Vice Chancellor for Research (OVCR) and the Athletic Department. Funding for this award has been provided through Big 10 Athletic Media Revenue (136-AAI3375). The authors also acknowledge the U.S. Office of Naval Research funding under the PANTHER award N00014-21-1-2044 through Dr. Timothy Bentley.

\item \textbf{Competing interests:} The authors have no competing interests to declare that are relevant to the content of this article.

\item \textbf{Author Contributions:} All authors contributed to the study conception and design. The experiments were performed by Anu Tripathi, Zhiren Zhu, Furkan Camci, Sheila Turcsanyi, Alison Brooks, Jeneel Pravin Kachhadiya, and Mauricio Araiza Canizales. 
Yang Wan, Zhiren Zhu, and Haneesh Kesari developed the codes for the A3G1 algorithm and continuous wavelet transform, and the adaptive filtering method was developed by Anu Tripathi with input from Christian Franck, Haneesh Kesari, and Rika Carlsen. The data analysis was performed by Anu Tripathi. The original draft was prepared by Anu Tripathi and revised by Rika Carlsen. All authors reviewed the manuscript and approved the final manuscript.
\end{itemize}

\noindent

\bigskip

\bibliography{sn-bibliography}

\end{document}